\documentstyle[12pt,aasms4,epsfig]{article}

\def\power#1{\mbox{$\times10^{#1}\ $}}

\newcommand{\msun}{M$_\odot$ }
\newcommand{\msyr}{M$_\odot$.yr$^{-1}$ }

\newcommand{\zaa}{A\&A~}
\newcommand{\zaas}{A\&AS~}
\newcommand{\zapj}{ApJ~}

\newcommand{\znp}{Nucl. Phys. }
\newcommand{\zprl}{Phys. Rev. Lett.}

\newcommand{\zpr}{Phys. Rev.}

\newcommand{\nam}{$^{21}$Na }
\newcommand{\na}{$^{22}$Na }
\newcommand{\naa}{$^{23}$Na }
\newcommand{\netonaa}{$^{20}$Ne(p, $\gamma$)$^{21}$Na($\beta^+$)$^{21}$Ne(p,
$\gamma$)$^{22}$Na($\beta^+$)$^{22}$Ne(p, $\gamma$)$^{23}$Na }
\newcommand{\neo}{$^{20}$Ne }
\newcommand{\nee}{$^{21}$Ne }
\newcommand{\neee}{$^{22}$Ne }
\newcommand{\nepgnam}{$^{20}$Ne(p, $\gamma$)$^{21}$Na }
\newcommand{\nepgna}{$^{21}$Ne(p, $\gamma$)$^{22}$Na }
\newcommand{\nepgnaa}{$^{22}$Ne(p, $\gamma$)$^{23}$Na }
\newcommand{\mgm}{$^{23}$Mg }
\newcommand{\napgmg}{$^{22}$Na(p, $\gamma$)$^{23}$Mg }
\newcommand{\mgbna}{$^{23}$Mg($\beta^+$)${}^{23}$Na }
\newcommand{\mgbnaa}{$^{22}$Mg($\beta^+$)${}^{22}$Na }
\newcommand{\nabne}{$^{21}$Na($\beta^+$)${}^{21}$Ne }
\newcommand{\nampgmg}{$^{21}$Na(p, $\gamma$)$^{22}$Mg }

\newcommand{\natomg}{$^{23}$Na(p,$\gamma$)$^{24}$Mg }
\newcommand{\napane}{$^{23}$Na(p,$\alpha$)$^{20}$Ne }
\newcommand{\altosi}{$^{25}$Al(p,$\gamma$)$^{26}$Si }
\newcommand{\algtosi}{$^{26}$Al$^{g}$(p,$\gamma$)$^{27}$Si }
\newcommand{\almtosi}{$^{26}$Al$^{m}$(p,$\gamma$)$^{27}$Si }
\newcommand{\sitop}{$^{26}$Si(p,$\gamma$)$^{27}$P }

\newcommand{\mgtoal}{$^{26}$Mg(p,$\gamma$)$^{27}$Al }
\newcommand{\mgpgal}{$^{23}$Mg(p,$\gamma$)$^{24}$Al }
\newcommand{\alsmall}{$^{23}$Al(p,$\gamma$)$^{24}$Si }

\newcommand{\gap}{\mathrel{ \rlap{\raise.5ex\hbox{$>$}}
                      {\lower.5ex\hbox{$\sim$}}  } }
\newcommand{\lap}{\mathrel{ \rlap{\raise.5ex\hbox{$<$}}
		    {\lower.5ex\hbox{$\sim$}}  } }

\newcommand{\al}{$^{26}$Al }

\newcommand{\all}{$^{27}$Al }
\newcommand{\mg}{$^{24}$Mg }
\newcommand{\mgg}{$^{25}$Mg }
\newcommand{\mggg}{$^{26}$Mg }

\begin{document}

\title{Nuclear uncertainties in the NeNa-MgAl cycles and production of
       ${}^{22}$Na and ${}^{26}$Al during nova outbursts}

 \author{Jordi Jos\'e} 
 \affil{Departament de F\'{\i}sica i Enginyeria Nuclear (UPC), Av.
 V\'{\i}ctor Balaguer, s/n, E-08800 Vilanova i la Geltr\'u (Barcelona), SPAIN}
 \affil{Institut d'Estudis Espacials de Catalunya (IEEC), 
 Edifici Nexus-201, C/ Gran Capit\`a 2-4, E-08034 Barcelona, SPAIN}
 \author{Alain Coc}
 \affil{Centre de Spectrom\'etrie Nucl\'eaire et de Spectrom\'etrie de
 Masse, IN2P3-CNRS, B\^at. 104, F-91405 Orsay Campus, FRANCE}
 \and
 \author{Margarita Hernanz}
 \affil{Institut d'Estudis Espacials de Catalunya (IEEC), 
 CSIC Research Unit,
 Edifici Nexus-201, C/ Gran Capit\`a 2-4, E-08034 Barcelona, SPAIN}

\received{}
\accepted{}

\slugcomment{\underline{Submitted to}: \zapj ~~~~\underline{Version}:
\today}

\begin{abstract}

Classical novae eject significant amounts of nuclear processed material into 
the interstellar medium. Among the isotopes synthesized during such explosions, 
two radioactive nuclei deserve a particular attention: ${}^{22}$Na and 
${}^{26}$Al. In this paper, we investigate the nuclear paths leading to 
${}^{22}$Na and ${}^{26}$Al production during nova outbursts by means of an 
implicit, hydrodynamic code that follows the course of the thermonuclear
runaway from the onset of accretion up to the ejection stage. New evolutionary 
sequences of ONe novae have been computed, using updated nuclear reaction rates 
relevant to ${}^{22}$Na and ${}^{26}$Al production. Special attention is 
focused on the role played by nuclear uncertainties within the NeNa and MgAl 
cycles in the synthesis of such radioactive species. From the series of 
hydrodynamic models, which assume upper, recommended or lower estimates of the 
reaction rates, we derive limits on the production of both ${}^{22}$Na and 
${}^{26}$Al. We outline a list of nuclear reactions which deserve new 
experimental investigations in order to reduce the wide dispersion 
introduced by nuclear uncertainties in the ${}^{22}$Na and ${}^{26}$Al yields.

\end{abstract}


\section{Introduction}

The thermonuclear runaway model has been successful in reproducing the 
{\it gross} features of nova outbursts. According to this widely accepted 
scenario, classical novae are produced by thermonuclear runaways (hereafter, 
TNRs) that take place in the white dwarf component of a close binary system. 
The large, main sequence companion overfills its Roche lobe, providing matter 
outflows through the inner Lagrangian point that lead to the formation of an 
accretion disk around the white dwarf. A fraction of this H-rich matter lost 
by the companion ultimately ends up on top of the white dwarf, where it is 
gradually compressed as accretion goes on. The piling up of matter heats the 
envelope up to the point when ignition conditions to drive a TNR are reached. 

An extended set of hydrodynamic computations of classical nova outbursts has 
been performed during the last 25 years (see Starrfield et al. 1972, for
the first hydrodynamic study of the TNR model, and the recent papers by 
Kovetz \& Prialnik 1997, Starrfield et al. 1998, Jos\'e \& Hernanz 1998, 
and references therein), for a wide 
range of white dwarf masses and initial luminosities, mass accretion rates 
and initial chemical compositions. From the nucleosynthesis viewpoint, these 
computations have been able to identify two types of nova outbursts, those 
occurring in CO or in ONe white dwarfs. The latter ones have provided a 
framework for the origin of the high concentrations of Ne and more massive 
isotopes found in the spectra of some well-observed novae, such as 
V693 CrA 1981, V1370 Aql 1982, QU Vul 1984 No. 2, V838 Her 1991 or V1974 Cyg 
1992 (Livio \& Truran 1994; Gehrz et al. 1998). 
Among the isotopes synthesized in these so-called 
Ne (or ONe) novae, two radioactive species have raised a particular 
astrophysical interest: ${}^{22}$Na and ${}^{26}$Al.

The potential role of \na for diagnosis of nova outbursts was first suggested 
by Clayton \& Hoyle (1974). It decays to a short-lived excited state of 
${}^{22}$Ne (with a lifetime of $\tau = 3.75$ yr), which de-excites to
its ground state by emitting a $\gamma$-ray 
photon of $1.275$ MeV. Through this mechanism, nearby ONe novae within a few 
kiloparsecs from the Sun may provide detectable $\gamma$-ray fluxes. Several 
experimental verifications of this $\gamma$-ray emission at 1.275 MeV from 
nearby novae have been attempted in the last twenty years, using balloon-borne 
experiments (Leventhal et al. 1977), and detectors on-board satellites such as 
HEAO-3 (Mahoney et al. 1982), SMM (Leising et al. 1988), and CGRO (Leising 
1993; Iyudin et al. 1995), from which upper limits on the ejected ${}^{22}$Na 
have been derived.  In particular, the observations performed with the COMPTEL 
experiment on-board CGRO of five recent Ne-type novae (Nova Her 1991, Nova Sgr 
1991, Nova Sct 1991, Nova Pup 1991 and Nova Cyg 1992. Iyudin et al. 1995), as 
well as observations of {\it standard} CO novae, have led to an upper limit of
$3.7 \times 10^{-8}$ \msun for the ${}^{22}$Na mass ejected by any nova in the 
Galactic disk. This restrictive limit has posed some constraints on 
pre-existing theoretical models of classical nova explosions. 

${}^{26}$Al is another unstable nucleus, with a lifetime of $\tau = 1.04 \times 
10^6$ years, that decays from ground state to the first excited level of \mggg, 
which in turn de-excites to its ground state by emitting a gamma-ray photon 
of 1.809 MeV.  This characteristic gamma-ray signature, first detected in the 
Galatic Center by the HEAO-3 satellite (Mahoney et al. 1982, 1984), has been 
confirmed by other space missions like SMM (\cite{Sha85}) and by several 
balloon-borne experiments.  The most recent measurements made with COMPTEL have 
provided a map of the 1.809 MeV emission in the Galaxy (Diehl et al. 1995, 
1997; Prantzos \& Diehl 1996). 
The inferred $1-3$ \msun of Galactic ${}^{26}$Al are, 
according to the observed distribution, mainly attributed to young progenitors, 
such as massive AGB stars, type II supernovae and Wolf-Rayet stars. 
More recent analyses of the 1.809 MeV COMPTEL map reveal a correlation between
this map and the COBE/DMR maps, tracing free-free emission, thus confirming 
that the main contributors
to the Galactic ${}^{26}$Al are massive stars (Kn\"odlseder 1997).
However, a 
potential  contribution from novae or low-mass AGB stars cannot be ruled out  
(see Jos\'e, Hernanz \& Coc 1997, for a  recent analysis of ${}^{26}$Al 
production in classical novae). 

First estimates of the ${}^{22}$Na and ${}^{26}$Al production in novae were 
performed by different groups, using simplified one-zone models with 
representative temperature and density profiles. Hillebrandt \& Thielemann 
(1982) and Wiescher et al. (1986) suggested already that classical novae might 
produce significant amounts of ${}^{26}$Al, not enough to represent major 
Galactic sources, but relevant to account for the observed isotopic anomalies 
found in some meteorites. New parametrized calculations by Weiss \& Truran 
(1990) and Nofar, Shaviv \& Starrfield (1991), 
revealed that nova envelopes previously enhanced in heavy elements 
(from Ne to Mg) produce large amounts of ${}^{22}$Na and ${}^{26}$Al. Since 
this metal enrichment is expected to result from dredge-up of core material, 
Weiss \& Truran suggested that massive ONeMg white dwarfs (the ones attaining 
the highest peak temperatures and, therefore, the most efficient dregde-up) are 
likely to provide the largest abundances of both ${}^{22}$Na and ${}^{26}$Al in 
the ejecta. Politano et al. (1995) revisited this scenario using hydrodynamic 
computations. They reported on a strong anticorrelation between ${}^{22}$Na and 
${}^{26}$Al production: novae that produce the largest amounts 
of ${}^{22}$Na (i.e., massive white dwarfs) are not the same as those 
accounting for the largest yields of ${}^{26}$Al (i.e., low-mass white dwarfs). 
The obtained ${}^{22}$Na yields range between $5 \times 10^{-5}$ and $5 \times 
10^{-3}$, by mass. Assuming that the whole envelope ($\sim 10^{-4}-10^{-5}$ 
\msun) is ejected during the outburst, they  concluded that
nearby ONeMg novae with $M_{wd} \geq 1.25$ \msun should produce 
detectable ${}^{22}$Na $\gamma$-rays for CGRO, a prediction not confirmed so 
far (Iyudin et al. 1995). Their results showed also a significant production 
of ${}^{26}$Al (i.e., $(19.6-7.5) \times 10^{-3}$, by mass, corresponding to 
ONeMg white dwarfs with masses between $1.0-1.35$ \msun, respectively), which 
could account for a major fraction of the Galactic ${}^{26}$Al. 

Recent hydrodynamic computations of ONe novae (Jos\'e, Hernanz \& Coc 1997; 
Jos\'e \& Hernanz 1997, 1998) using  updated initial compositions and nuclear 
reaction rates, have led to a significant reduction of both ${}^{26}$Al and 
${}^{22}$Na ejected during nova outbursts.  In particular, a mean mass fraction 
of $1 \times 10^{-4}$ of ${}^{22}$Na is found in the 1.25 \msun ONe Model (with 
$M_{ejec}$(${}^{22}$Na)=$ 1.3 \times 10^{-9} $ \msun), whereas a maximum value 
of $6 \times 10^{-4}$ results from the 1.35 \msun ONe Model 
($M_{ejec}$(${}^{22}$Na)=$ 2.6 \times 10^{-9} $ \msun). The corresponding peak 
fluxes in the 1.275 MeV ${}^{22}$Na line, below $10^{-5}$ photons s$^{-1}$ 
cm$^{-2}$ for novae at 1 kpc, turn out to be too low to be detected with OSSE 
or COMPTEL but represent potential targets for the nearby future INTEGRAL 
mission (Hernanz et al. 1997; G\'omez-Gomar et al. 1998). Concerning 
${}^{26}$Al, yields ranging from $2 \times 10^{-3}$ to $2 \times 10^{-4}$ by 
mass have been obtained in a series of ONe nova models with masses between 
$1.15-1.35$ \msun. Contribution of novae to the Galactic ${}^{26}$Al is limited 
to $0.1-0.4$ \msun  (Jos\'e, Hernanz \& Coc 1997). Nevertheless, a larger 
contribution cannot be ruled out if the (uncertain) lower limit for ONe white 
dwarfs is reduced down to 1.0 \msun (Jos\'e \& Hernanz 1998).

Other hydrodynamic computations performed by Starrfield et al. (1997, 1998), 
using also updated nuclear reaction rates and opacities, have modified their 
previous estimates (Politano et al. 1995). The expected abundance of 
${}^{22}$Na in the ejecta has risen up to $(2-3) \times 10^{-3}$, by mass, 
when 1.25 \msun ONeMg white dwarfs are adopted, high enough to be detected by 
CGRO, provided that all the accreted envelope ($3 \times 10^{-5}$ \msun) 
is ejected. On the other hand, the improved input physics translates into a 
factor of 10 reduction  on the synthesis of ${}^{26}$Al, in better agreement 
with the analysis of the 1.809 MeV emission map provided by COMPTEL, and also 
with the results previously reported by Jos\'e, Hernanz \& Coc (1997). 

Whereas the agreement between the different groups on the expected contribution 
of classical novae to the Galactic ${}^{26}$Al has significantly increased, 
there remains some discrepancy concerning the amount of ${}^{22}$Na present in 
the ejecta. Since the synthesis of both ${}^{22}$Na and ${}^{26}$Al is very 
dependent on the adopted nuclear reaction rates, the large uncertainties 
present in some key reactions of both NeNa and MgAl cycles (Coc et al. 1995; 
Prantzos \& Diehl 1996), may significantly modify the expected yields. In 
particular, the study of the influence of a given reaction rate on \na 
production is not trivial due to the two possible modes of formation and their 
interplay with convection (Coc et al. 1995). Moreover, during a nova outburst, 
thermodynamic conditions change on a short timescale so that nuclear reactions 
are never close to equilibrium. Due to
obvious experimental difficulties, reaction rates involving short-lived
radioactive nuclei of the NeNa-MgAl group are in general 
poorly known and the associated uncertainties may reach many orders of 
magnitude. It is thus important to know how the yields of important isotopes 
like $^{22}$Na and $^{26}$Al are affected by those uncertainties. 
In this paper, series of hydrodynamic nova models have been computed 
assuming upper, recommended and lower estimates of the reaction rates, from 
which limits on 
the production of both ${}^{22}$Na and ${}^{26}$Al are derived. The present 
analysis is focused on capture rates on radioactive nuclei, and for a 
temperature domain in the range $T_8 \simeq$ 0.5 -- 3.5. 
We refer to the NACRE 
compilation (Angulo et al. 1998) for a more general discussion concerning 
capture rates on stable isotopes, and a wider temperature range. For 
convenience, we use the terms NeNa cycle and MgAl cycle to denote the reactions 
involved in $^{22}$Na and $^{26}$Al formation. However, because of the high 
$^{23}$Na(p,$\gamma$) and $^{27}$Al(p,$\gamma$) rates, they cannot be 
considered as genuine {\em cycles}. 

In Section 2, we outline some details of the method of computation and input 
physics. A detailed analysis of the synthesis of ${}^{22}$Na and ${}^{26}$Al 
in classical novae, together with the study of the role played by specific 
nuclear reactions of the NeNa-MgAl cycles, is given in Sections 3 \& 4. 
Constraints on the production of ${}^{22}$Na and ${}^{26}$Al assuming lower, 
recommended and upper rates for some key reactions are derived in Section 5. 
The most relevant 
conclusions of this paper are summarized in Section 6. A detailed Appendix 
focused on the nuclear physics aspects of the reaction rates within the NeNa 
and MgAl cycles follows.

\section{Model and Input physics}

Evolutionary sequences of nova outbursts have been calculated by means of an 
updated version of the code SHIVA (see Jos\'e 1996; Jos\'e \& Hernanz 1998), a 
one-dimensional, implicit, hydrodynamical code in lagrangian formulation, that 
follows the course of the outburst from the onset of accretion up to the 
expansion and ejection stages. The code solves the standard set of differential 
equations for hydrodynamical evolution: conservation of mass, momentum and 
energy, energy transport by radiation and convection, plus the definition of 
the lagrangian velocity, including a time-dependent formalism for convective 
transport whenever the characteristic convective timescale becomes larger than 
the integration time step (Wood 1974). Partial mixing between adjacent 
convective shells is treated by means of a diffusion equation (see 
Prialnik, Shara \& Shaviv 1979, for the formalism). 
The code is linked to a reaction network, which follows the 
detailed evolution of 100 nuclei, ranging from $^{1}$H to $^{40}$Ca, through 
370 nuclear reactions, with updated rates, and screening factors from Graboske 
et al. (1973) and DeWitt, Graboske \& Cooper (1973). As suggested by Politano 
et al. (1995), the matter transferred from the companion is assumed to be 
solar-like, and is mixed in a given fraction with the outermost shells of the 
underlying core by means of an unknown mechanism (either shear mixing, 
diffusion or a convective multidimensional process).  
 The composition of the underlying core has been taken from 
recent detailed evolutionary models in the case of ONe white dwarfs, which are 
the main contributors to \na and \al synthesis. These stars are made basically 
of $^{16}$O and $^{20}$Ne (\cite{Dom93}; \cite{Rit96}), with smaller traces of 
${}^{23}$Na, ${}^{24,25}$Mg, ${}^{27}$Al and other species. This issue plays a 
crucial role in the resulting nucleosynthesis, and should be taken into account 
in order to compare results obtained by different groups. In particular, the 
ONeMg models computed by Starrfield et al. (1998) have an initial composition 
of the white dwarf core based on old nucleosynthesis calculations of C-burning 
from Arnett \& Truran (1969), which is richer in ${}^{24}$Mg and ${}^{20}$Ne 
than the one adopted in this paper (see Table 1). 

\section{Synthesis of ${}^{22}$Na in Classical Novae}

The high temperatures attained during ONe nova outbursts  (with peak values 
within $2 - 3.5 \times 10^8$ K), allow a noticeable nuclear activity in 
the NeNa and MgAl cycles, which results on a significant production of species 
of astrophysical interest, such as ${}^{22}$Na and ${}^{26}$Al. In this 
Section, we will describe the main mechanisms of \na synthesis, through a 
detailed analysis of a 1.25 \msun, ONe white dwarf,
 which accretes solar-like matter at a rate $\dot M = 2 \times 
10^{-10}$ \msyr, assuming a 50\% degree of mixing with the ONe core 
 (Model ONe5, in Jos\'e \& Hernanz 1998).  Snapshots of 
the evolution of several isotopes relevant to ${}^{22}$Na synthesis (i.e., 
${}^{20,21,22}$Ne, ${}^{21,22,23}$Na and ${}^{22,23}$Mg) are shown in 
Figure 1. 

\subsection{Main nuclear reactions involved in ${}^{22}$Na production}

At the onset of accretion, the evolution of \na is mainly dominated by the 
chain of nuclear reactions \netonaa (i.e., the cold mode of the NeNa cycle.
See Fig. 2). 
When the temperature at the burning shell reaches $T_{bs} =$ 
5\power{7} K (Fig. 1, panel 1), the main nuclear reaction of the NeNa cycle 
is \nepgna, which significantly reduces the amount of \nee. In fact,  when 
$T_{bs} =$ 7\power{7} K (panel 2), the amount of \nee is already too small 
to maintain the main mechanism for \na synthesis. Therefore, \na will begin to 
decrease near the burning shell due to proton captures, following the rise of 
temperature toward peak value (panels 3 \& 4).
 
At $T_{bs} =$ 1\power{8} K (panel 3), the amount of \nee has already 
decreased below $10^{-6}$ by mass, except at the outer envelope, where some 
\nee is synthesized from the $\beta^+$-decay of \nam (previously built up as 
\nepgnam). \nam and \mgm increase due to \nepgnam and \napgmg, respectively.
Destruction of \na through (p, $\gamma$) reactions goes on but, due to 
convection, \na shows a nearly flat profile throughout the envelope. 

When $T_{bs} =$ 2\power{8} K (panel 4), there is a dramatic decline in \naa 
(due to (p,$\gamma$) and (p,$\alpha$) reactions) and in the \na abundance (by 
\napgmg).  \neee has slightly decreased. Also noticeable is the increase of 
\nam (since \nepgnam dominates destruction from both \nabne and \nampgmg),  
which plays a crucial role in the synthesis of \na at the late stages of the 
outburst. Both ${}^{22,23}$Mg increase due to proton captures on 
${}^{21,22}$Na respectively.

33 seconds later, the temperature at the burning shell attains its peak value, 
$T_{bs, max} = 2.44 \times 10^8$ K (panel 5). The amount of \nam is maintained 
by a quasi-equilibrium between \nampgmg and \nepgnam. The mean amount of \na 
increases due to \mgbnaa, previously transported by convection to the outer, 
cooler layers of the envelope. The amount of \naa decreases due to both 
(p,$\gamma$) and (p,$\alpha$) reactions, which dominate \nepgnaa as well as 
\mgbna.  Since the peak temperature achieved in the burning shell is not 
extremely high, \neo remains nearly unchanged (the typical temperature for 
\neo burning exceeds $\sim 4  \times 10^8$ K). With respect to the other neon 
isotopes, \nee increases (mainly due to the $\beta^+$-decay of \nam at the 
outer shells), whereas \neee is destroyed by \nepgnaa. The two magnesium 
isotopes ${}^{22,23}$Mg increase as a result of proton captures onto 
${}^{21,22}$Na.

 Shortly after, due to the sudden release of energy from the short-lived 
$\beta^+$-unstable 
nuclei ${}^{13}$N, ${}^{14,15}$O, and ${}^{17}$F, the envelope begins to 
expand. The role played by (p,$\gamma$) and (p,$\alpha$) reactions is 
therefore reduced following the drop in temperature, whereas $\beta^+$-decays 
progressively dominate the evolution. The abundances of ${}^{21}$Na and 
${}^{22,23}$Mg decrease as a result of such $\beta^+$-decays, which in turn 
increase the amount of ${}^{21}$Ne and ${}^{22,23}$Na (panels 6 \& 7).

At the final stages of the outburst (panel 8),  as the envelope expands 
and cools down, most of the remaining nuclear activity in the NeNa cycle is due 
to $\beta^+$-decays, such as  \nabne, \mgbnaa or \mgbna. The resulting mean 
abundance of \na in the ejected shells of this $1.25$ \msun ONe Model is 
X(\na) = 9.6\power{-5} by mass, which corresponds to 1.3\power{-9} \msun of 
\na ejected into the interstellar medium. Other species of the NeNa cycle 
present in the ejecta are \neo (X(\neo) = 0.18, a mass fraction slightly
higher than its initial value because of the operation of the
\napane reaction. See Table 1),  ${}^{21,22}$Ne (X(\nee) = 3.5\power{-5}, 
and X(\neee) = 1.0\power{-3}), and \naa (with X(\naa) = 1.4\power{-3}). 

\subsection{Effect of the reaction rates on the synthesis of ${}^{22}$Na}

From the abovementioned nuclear physics viewpoint, 
the synthesis of \na is mainly controlled 
by four reactions: \nepgnam, \nampgmg, \nepgna and \napgmg (Fig. 2), for which 
recent updates to the reaction rates are available. Despite energy production 
during nova outbursts is not very dependent on the specific prescriptions 
adopted for such rates (see Section 5.1), they play a crucial role on the 
accompanying nucleosynthesis, since changes in the reaction paths are expected. 
Several test models have been computed to analyse the role played by
different nuclear reactions on the synthesis of ${}^{22}$Na (as well as
${}^{26}$Al). Main results concerning  ${}^{22}$Na and ${}^{26}$Al production,
compared with the abundances found in Jos\'e \& Hernanz (1998) with
previous prescriptions, are summarized in Table 2. 

%***20Ne(p,g)
\nepgnam is the slowest proton capture reaction on any stable neon and sodium
isotope. According to the adopted composition for the ONe core (Ritossa,
Garc\'\i a--Berro \& Iben 1996), ${}^{20}$Ne is the most abundant Ne-Na
isotope. Therefore, \nepgnam limits \na production in classical novae. The
adopted \nepgnam rate is the Caughlan \& Fowler (1988) one, which is based on
Rolfs et al. (1975) data.
In the considered domain of temperature, it is well known.
The estimated uncertainty is of a factor of $\simeq$3 (defined as the ratio
between high and low estimates within the domain of temperature considered), 
according to the 
new compilation of nuclear reaction rates (Angulo et al. 1998), but 
more likely a factor of $\simeq$1.5 only,  as it is derived from 
short range standard extrapolations of experimental data (Rolfs et al. 1975).
Therefore, nuclear uncertainties associated with \nepgnam should play no 
significant role on \na production.

%***21Ne(p,g)
Concerning the \nepgna rate, no significant effect of the reduction of the 
contribution of the $E_x=6.384$ MeV level (G\"orres et al. 1983) 
with respect to Caughlan \& Fowler (1988) has been found on \na production 
(tested with a 1.25 \msun ONe white dwarf model). 
According to Angulo et al. (1998), other nuclear 
uncertainties scarcely affect the \nepgna rate at moderate temperatures. On 
the contrary, much more uncertain is the \nepgnaa rate, poorly known in the 
range of temperatures of interest for nova outbursts (Angulo et al. 1998). 
Nevertheless, due to the low initial ${}^{22}$Ne abundance with respect to 
${}^{23}$Na, and also to the negligible $^{22}$Na($\beta^+$)${}^{22}$Ne decay 
($\tau =$ 3.75 yr) during the TNR, the nuclear uncertainties affecting the 
\nepgnaa rate are not relevant for \na production in novae.

%***22Na(p,g)
Recent experimental investigations of the ${}^{22}$Na(p,$\gamma$)${}^{23}$Mg 
reaction (Seuthe et al. 1990; Schmidt et al. 1995; Stegm\"uller et al. 1996) 
have provided a firmer basis for the determination of its rate. Since the new 
rate is lower than the old Caughlan \& Fowler's (1988) analytic fit (i.e., one 
order of magnitude for $T_8 > 1$), destruction of \na by means of 
(p,$\gamma$) reactions is reduced. In a test model, consisting of a 1.25 \msun 
ONe white dwarf (Model ONe5, Jos\'e \& Hernanz 1998), the mean abundance of 
\na in the ejecta increases by a factor of $\sim$ 3 when the new \napgmg rate 
(Stegm\"uller et al. 1996), instead of the CF88 one, is adopted (see Table 2). 
It is worth 
noticing that significant nuclear uncertainties affect this rate (a factor
ranging from 3 to 6 for $T_8 > 1$), which turn out to be 
crucial in order to derive ranges of \na production during nova outbursts. 
According to \cite{Ste96}, this uncertainty is mainly due to the possible 
effect of a resonance at $E_p$ = 225~keV. Below $T_8$=1, the uncertainty 
reaches three orders of magnitude but the rate remains small enough to 
prevent destruction of $^{22}$Na. 

%***21Na(p,g)
Three test models have been computed to analyze the role played by \nampgmg,
in view of the nuclear uncertainties present in this rate, in particular the
estimated strength of the first $E_x=5.714$ MeV level above the proton
threshold (see Appendix, section A.1). 
Calculations assume ONe white dwarfs of masses 1.15,
1.25 and 1.35 \msun, and the same input physics than Models ONe3, ONe5 and ONe6
described in Jos\'e \& Hernanz (1998), but reducing the \nampgmg rate given by
CF88 by a factor of 100 (similar, for novae, to the lower rate given in the
Appendix). It results in a significant increase in the \na abundances present
in the ejecta (a factor of $\sim$ 2 to 3, in the 1.15 and 1.25 \msun 
Models, and a factor of $\sim$ 1.2, in the 1.35 \msun Model. Table 2), 
as compared with the values found
with the standard CF88 rate. This effect can be interpreted as follows: 
when the 
${}^{21}$Na(p,$\gamma$)${}^{22}$Mg($\beta^+$)${}^{22}$Na rate is reduced
by a factor of 100, the alternative path,
${}^{21}$Na($\beta^+$)${}^{21}$Ne(p,$\gamma$)${}^{22}$Na, is favored.
In this case, \na production is delayed to a time when the envelope
is already expanding and cooling down (contrary to the case when the
higher \nampgmg rate is adopted). 
As a result, a major fraction of \na survives. 
This, in turn, explains the
lower effect found in the 1.35 \msun Model, caused by the higher temperatures
achieved in the envelope (with $T_{peak} = 3.2 \times 10^8$ K) which remain
high enough at the time when \na is synthesized.
 One should note that this effect (increase in the \na yield when
the ${}^{21}$Na(p,$\gamma$)${}^{22}$Mg rate is reduced) was not foreseen 
and stresses the importance of full hydrodynamical calculations.  

Other reactions that may be involved in the synthesis of \na  (and ${}^{23}$Na) 
are ${}^{22}$Mg(p,$\gamma$)${}^{23}$Al,  ${}^{23}$Al(p,$\gamma$)${}^{24}$Si, or 
${}^{23}$Mg(p,$\gamma$)${}^{24}$Al. However, no significant effect on the \na 
production is found when using updated rates for such reactions (see Appendix
 and Table 2), because of the limited nuclear flow they conduct 
for nova conditions.

According to this analysis, full evolutionary sequences of nova 
outbursts, from the onset of accretion up to the ejection stage, have been  
performed taking into account the low, recommended and high estimates to the 
\nampgmg, \napgmg  and \nepgna rates (as well as the new 
${}^{23}$Mg(p,$\gamma$)${}^{24}$Al rate derived by Kubono, Kajino \& Kato 
1995), in order to derive a range of \na yields, resulting purely from nuclear 
physics uncertainties. Results are summarized in Section 5.

\section{Synthesis of ${}^{26}$Al in Classical Novae}

In this Section, we will analyze the nuclear paths leading to ${}^{26}$Al 
synthesis.  As for ${}^{22}$Na production, we will describe the course of the 
1.25 \msun, ONe nova model (Jos\'e \& Hernanz 1998, Model ONe5). Snapshots of 
the evolution of several isotopes relevant to ${}^{26}$Al synthesis (i.e., 
${}^{24,25,26}$Mg, ${}^{25,27}$Al, ${}^{26,27}$Si and the ground and 
isomeric states for ${}^{26}$Al, herefater ${}^{26}$Al$^{g}$ 
and ${}^{26}$Al$^{m}$) are shown in Figure 3. 

\subsection{Main nuclear reactions involved in ${}^{26}$Al production}

Nucleosynthesis of \al is complicated by the presence of a short lived 
($\tau_{1/2}$ = 6.3~s) spin isomer.
At low temperatures ($T_8\lap4$), the 
$^{26}$Al ground and isomeric states do not reach thermal equilibrium and must 
be treated as two separate isotopes (Ward \& Fowler 1980, see also 
Coc \& Porquet 1998).

The nuclear activity in the MgAl cycle at the early phases of the TNR, when
the temperature at the burning shell is $T_{bs} = 5 \times 10^7$ K, is
dominated by ${}^{25}$Mg(p,$\gamma$), which leads to both ${}^{26}$Al ground
and isomeric states.  A significant amount of ${}^{26}$Al$^g$ is already 
synthesized at such temperatures (Figure 3, panel 1). Another aluminum 
isotope, 
${}^{27}$Al, is slightly enhanced with respect to its initial abundance by 
means of ${}^{26}$Al$^m$($\beta^+$)${}^{26}$Mg(p,$\gamma$)${}^{27}$Al. A 
similar trend is found when $T_{bs} = 7 \times 10^7$ K (panel 2): both 
${}^{26}$Al$^g$ and ${}^{27}$Al continue to rise (specially the first one, 
which increases by nearly a factor of 10 from the abundance shown in panel 1).

At $T_{bs} = 10^8$ K (panel 3), the evolution in the MgAl cycle is dominated by 
${}^{24}$Mg(p,$\gamma$)${}^{25}$Al($\beta^+$)${}^{25}$Mg,  which in turn 
accounts for a noticeable production of the $\beta^+$-unstable nuclei 
${}^{25}$Al, and also for the increase in the mean ${}^{26}$Al$^g$ abundance 
(by means of ${}^{25}$Mg(p,$\gamma$)${}^{26}$Al$^{g,m}$). A nearly flat profile 
of ${}^{26}$Al$^g$ results from convective mixing, which extends already 
throughout the whole envelope.

A major change in the dominant nuclear path is found when $T_{bs} = 2 \times 
10^8$ K (panel 4). The temperature attained near the burning shell is high 
enough to allow (p,$\gamma$) reactions to proceed efficiently. In particular, 
the abundance of ${}^{24}$Mg is reduced by a factor of $\sim$ 1000. The isomer 
${}^{26}$Al$^m$ exceeds already $10^{-3}$ by mass for most of the envelope. 
A significant fraction of ${}^{25}$Al is transformed through proton captures 
into ${}^{26}$Si, which will increase the abundance of ${}^{26}$Al$^m$, and in 
turn that of ${}^{27}$Al in the late phases of the TNR. The final 
${}^{27}$Al/${}^{26}$Al$^g$ ratio will reflect a competition between two 
different paths: ${}^{24}$Mg (fed by ${}^{23}$Na(p,$\gamma$)${}^{24}$Mg) is 
transformed by proton captures into ${}^{25}$Al, which can either decay into 
${}^{25}$Mg or capture another proton, leading to ${}^{26}$Si. Only the first 
channel accounts for ${}^{26}$Al$^g$ synthesis, whereas ${}^{27}$Al can be 
produced by both paths, following ${}^{25}$Mg(p,$\gamma$)${}^{26}$Al$^{g,m}$(p, 
$\gamma$)${}^{27}$Si($\beta^+$)${}^{27}$Al, or 
${}^{26}$Si($\beta^+$)${}^{26}$Al$^m$($\beta^+$)${}^{26}$Mg(p, 
$\gamma$)${}^{27}$Al.  Anyway, because of the large abundance of ${}^{25}$Mg
in the envelope at this stage, proton captures on the seed $^{25}$Mg
become the major source of $^{26}$Al$^g$.
Some leakage from the MgAl cycle due to 
${}^{26}$Al$^{g,m}$(p,$\gamma$)${}^{27}$Si is also obtained at this stage of 
the outburst.

When the burning shell attains its peak temperature ($T_{bs,max} = 2.44 \times 
10^8$ K. Panel 5), most of the MgAl isotopes show a significant reduction near 
the burning shell because of (p,$\gamma$) reactions. The dominant paths at this 
stage are ${}^{25}$Mg(p,$\gamma$)${}^{26}$Al$^g$(p,$\gamma$)${}^{27}$Si and 
${}^{27}$Al(p,$\gamma$)${}^{28}$Si, which also account for a significant 
leakage from the cycle (moreover, ${}^{26,27}$Si(p,$\gamma$) are much faster 
than the corresponding $\beta^+$-decays). Following the course of the outburst 
(panels 6 to 8), as the envelope expands and cools down, 
proton capture reactions are progressively reduced. Therefore, the late time 
evolution is mainly dominated by $\beta^+$-decays, such as 
${}^{26}$Si($\beta^+$)${}^{26}$Al$^m$, ${}^{27}$Si($\beta^+$)${}^{27}$Al, 
${}^{25}$Al($\beta^+$)${}^{25}$Mg and ${}^{26}$Al$^m$($\beta^+$)${}^{26}$Mg, 
which in turn, increase the final amounts of ${}^{25,26}$Mg and ${}^{27}$Al.  
The mean abundance of ${}^{26}$Al in the ejecta of this 1.25 \msun, ONe white 
dwarf model is X(${}^{26}$Al) = $5.4 \times 10^{-4}$ by mass, which translates 
into $7.6 \times 10^{-9}$ \msun of ${}^{26}$Al ejected into the interstellar 
medium. Other isotopes of the MgAl group present in the ejecta are ${}^{27}$Al 
(X(${}^{27}$Al) = $2 \times 10^{-3}$, half the initial abundance. See Table 
1), ${}^{25}$Mg ($2.4 \times 10^{-3}$), ${}^{26}$Mg ($2.8 \times 10^{-4}$), 
and ${}^{24}$Mg ($2 \times 10^{-4}$). 

\subsection{Effect of the reaction rates on the synthesis of ${}^{26}$Al}

The previous analysis has revealed that several isotopes should be considered 
as potential seeds for \al synthesis: 
\mg, \mgg and, to some extent, \naa and \neee. In this case, the number of 
nuclear reactions involved in the synthesis of \al is rather large (see Fig. 
2). Several test models of nova outbursts have been computed to analyze the 
role played by each one of the most relevant reactions. Main results are
summarized in Table 2.

%***23Na(p,g)
The $^{23}$Na(p,$\gamma)^{24}$Mg rate (in contrast with 
$^{23}$Na(p,$\alpha)^{20}$Ne) has been strongly modified by the 
introduction of the upper limit for the $E_r^{cm}$=0.138~MeV resonance strength 
obtained by G\"orres, Wiescher \& Rolfs (1989). When taking this upper limit, 
the rate is significantly increased up to $T_8\simeq$2 (G\"orres et al. 1989). 
Its effect has been tested by recomputing a  1.15 \msun, ONe white dwarf Model 
(i.e., Model ONe3. Jos\'e \& Hernanz 1998): an increase by a factor of $\sim$ 3 
of the final \mg yields results, which in turn leads to slightly larger mass 
fractions of both \mgg (factor 1.4) and \al (factor 1.3). Some \all is also 
overproduced (by a factor of 1.4) when this resonance is taken into account. 
Those effects are expected to be stronger for more massive white dwarfs. 

%=======mg24pg
At the temperatures attained in nova outbursts, $^{24}$Mg(p,$\gamma)^{25}$Al 
may become even faster than $^{12}$C(p,$\gamma)^{13}$N. The 
$^{24}$Mg(p,$\gamma)^{25}$Al rate is dominated by the presence of the 
$E_r^{cm}$ = 0.214~MeV, $J^\pi$= $1/2^+$ resonance and suffers little 
uncertainty. 

%====mg25pg====
$^{25}$Mg(p,$\gamma)^{26}$Al$^g$ is also of great importance since it is the
only channel that leads to $^{26}$Al in its ground state. One may expect that
its potential uncertainties are directly reflected in the ${}^{26}$Al$^g$
yields. The new rate is significantly smaller
by a factor of $\simeq$7 (\cite{Ili96}) than the \cite{CF88} one, 
below $T_8\simeq 1$, but this has no consequences in 
the nova domain. 
 \cite{Ili96} reanalyzed available transfer reaction data concerning 
 the lower lying resonances and concluded that the associated
uncertainties are relatively small (i.e., a factor of 1.5--2).
 For $T_8 > 1.5$, available 
direct measurements lead also to very small uncertainties 
(a factor of 1.5, according 
to \cite{Ili96}) and should not influence significatively \al production. 
Moreover, since $^{25}$Mg(p,$\gamma$) can lead either to $^{26}$Al$^g$ or 
$^{26}$Al$^m$ with a known branching ratio (Endt \& Rolfs 1987, 
Iliadis et al. 1996), the $^{25}$Mg(p,$\gamma$)$^{26}$Al$^m$ rate is known 
with a similar precision as the $^{25}$Mg(p,$\gamma$)$^{26}$Al$^g$ one.

%***26Al,g(p,g)
In nova nucleosynthesis, the most important resonance in 
$^{26}$Al$^g$(p,$\gamma)^{27}$Si is the $E_r^{cm}$ = 0.188~MeV, while 
uncertainties on the lower lying resonances have no influence (Coc et al. 
1995).  In our calculations, we use the only single direct measurement 
available (Vogelaar 1989) 
for this crucial resonance strength. Recently, the results of the proton 
transfer reaction that motivated this subsequent direct measurement have been 
published (Vogelaar et al. 1996) but the resonance strength deduced from this 
indirect measurement is strongly dependent on the adopted transferred angular  
momentum. Accordingly, we have checked the influence of such resonance on the 
resulting \al yields by reducing its strength by a factor of 
1/3, well within the 
range of values adopted by \cite{NACRE}. We obtain a significant increase by a 
factor of $\sim$ 2 in the final amount of \al (for a 1.15 \msun ONe white 
dwarf Model. Table 2), 
which strongly stresses the need of additional direct 
measurements to confirm the results found by Vogelaar (1989).

%***25Al(p,g)
${}^{25}$Al(p,$\gamma$)${}^{26}$Si plays an important role on \al synthesis, 
since it leads to the formation of the short-lived isomer (through 
${}^{26}$Si($\beta^+$)${}^{26}$Al$^m$), instead of the long-lived ground state. 
We have checked the role played by the 
${}^{25}$Al(p,$\gamma$)${}^{26}$Si rate, adopting the upper and lower 
estimates (see Appendix) provided by Coc et al. (1995). Results are
compared with those
obtained with the  ${}^{25}$Al(p,$\gamma$)${}^{26}$Si rate given by Wiescher 
et al. (1986).  Whereas the final amounts of \al and \all are very slightly 
enhanced when the lower rate is adopted (i.e., case A. Coc et al. 1995), a 
significant reduction of \al by a factor of $\sim 2$ is obtained for the upper 
one (case C). In the first case, ${}^{25}$Al($\beta^+$)${}^{25}$Mg becomes 
faster than ${}^{25}$Al(p,$\gamma$)${}^{26}$Si and, therefore, the final 
amount of \mgg (and in turn \al) increases (Table 2).  

%***26Al,m(p,g)
Large uncertainties affect the ${}^{26}$Al$^m$(p,$\gamma$)${}^{27}$Si rate. The 
existing Caughlan \& Fowler's (1988) prescription for this rate results from a 
Hauser-Feschbach calculation, a statistical approach that is not reliable at 
low temperatures, where the contribution of isolated resonances dominates. 
Hence, the rate may differ by several orders of magnitude from the theoretical 
one at the temperatures achieved during nova outbursts. However, no noticeable 
effect on ${}^{26}$Al production 
is found when the ${}^{26}$Al$^m$(p,$\gamma$)${}^{27}$Si rate, as given 
by CF88, is multiplied arbitrarily by a factor of 100, with the exception of a 
net reduction (factor $\sim$ 2) on the final amount of \mggg. 

%***26Mg(p,g)
Nuclear uncertainties significantly affect the $^{26}$Mg(p,$\gamma)^{27}$Al 
rate. Since the initial amount of $^{26}$Mg is of the same order of magnitude 
than that of $^{27}$Al, and it is also fed continuously by 
$^{26}$Al$^m$($\beta^+)^{26}$Mg ($\tau$ = 9.15~s), these uncertainties cannot 
be ignored as for $^{22}$Ne(p,$\gamma)^{23}$Na (see Section 3.2). They are 
specially important around $T_8 = 0.5$, where the rate remains uncertain by a 
factor of $\simeq 100$ (Champagne et al. 1990), due to the unknown strength of 
an hypothetical resonance at $E_R^{cm}$ = 90~keV. No influence on the final 
 ${}^{26,27}$Al yields results from the inclusion of the 
${}^{26}$Mg(p,$\gamma$)${}^{27}$Al recommended rate given by Champagne et al. 
(1990) (tested with a 1.15 \msun ONe white dwarf Model). 

%====al27pa====
With respect to CF88, the $^{27}$Al(p,$\gamma)^{28}$Si rate has not changed:
above $T_8$=1, uncertainties are limited to a factor of $\simeq $2 
(Angulo et al. 1998).  On the contrary, new experimental data for the 
$^{27}$Al(p,$\alpha)^{24}$Mg rate appeared shortly after the CF88 compilation: 
a direct measurement by \cite{Tim88} and a study of proton and alpha emission 
from $^{28}$Si levels performed by \cite{Cha88}. As a result, the rate is 
strongly reduced with respect to the Caughlan \& Fowler (1988) one by up to 
 4 orders of magnitude in the region of interest. As a consequence, the 
calculated \al yields were strongly reduced (Jos\'e, Hernanz \& Coc 1997).

%******Several rates
The new rates available for ${}^{23}$Al(p,$\gamma$)${}^{24}$Si 
(Schatz et al. 1997), 
${}^{26}$Si(p,$\gamma$)${}^{27}$P (Herndl et al. 1995), and 
${}^{23}$Mg(p,$\gamma$)${}^{24}$Al 
(Kubono, Kajino \& Kato 1995 and Herndl et al. 1998) have no 
significant effect on the resulting \al and \all yields, as compared with the 
results obtained by Jos\'e \& Hernanz (1998) with $1.25$ \msun ONe white 
dwarfs, using earlier prescriptions for these rates (i.e., van Wormer et al. 
1994, Wiescher et al. 1986, and also Wiescher et al. 
1986, respectively).

\section{Updated nuclear reaction rates, nuclear uncertainties and their
            effect on the synthesis of ${}^{22}$Na and ${}^{26}$Al}

Constraints on the synthesis of ${}^{22}$Na and ${}^{26}$Al during nova 
outbursts have been derived by means of hydrodynamic calculations involving 
ONe white dwarfs of masses 1.15, 1.25 and 1.35 \msun, assuming lower, 
recommended and upper estimates for several key reactions of the NeNa-MgAl 
cycles. 
Table 3 lists the set of reaction rates adopted for the calculation of 
minimum to maximum ${}^{22}$Na and ${}^{26}$Al production, compared with the 
rates adopted in Jos\'e \& Hernanz (1998). 
The specific choice adopted for these three sets is directly determined by the
range of nuclear uncertainties accompanying the rates. The corresponding
yields obtained with the different nuclear reaction networks, which are 
summarized in Tables 4 to 6, allow to derive error bars to the synthesis
of ${}^{22}$Na and ${}^{26}$Al. Since we have not included other sources of
uncertainty (convection, modelization of the explosion, ...)
 these error bars are only of nuclear physics origin. 

\subsection{Characteristics of the explosion}

The early stages of the outburst are mainly dominated by the CNO cycle (Jos\'e 
\& Hernanz 1998). Therefore, the update of the nuclear reaction rates of the 
NeNa-MgAl cycles has no influence on the characteristics of the accretion phase 
(for instance, the duration of the accretion phase, $t_{acc}$, or the mass of 
the accreted envelope, $\Delta M_{env}$).
 
Differences in the time evolution appear when the temperature near the burning 
shell reaches $T_{bs} \sim 10^8$ K. At this stage, energy generation by nuclear 
reactions involves a relevant contribution from the NeNa-MgAl cycles, together 
with the hot and cold modes of the CNO cycle.  The use of updated rates 
modifies several properties of the TNR, such as the time required for a 
temperature rise from $3 \times 10^7$ K up to $10^8$ K (hereafter, $t_{rise}$). 
For instance, a value of $t_{rise} = 6.8 \times 10^6$ s was found for Model 
ONe5 (Jos\'e \& Hernanz 1998), whereas a shorter time, $t_{rise}=4.3 \times 
10^6$ s, has been obtained in Model ONe125B, computed with updated 
(recommended) rates. 

The role played by the NeNa-MgAl cycles increases when 
the temperature at the location of the burning shell exceeds $2 \times 10^8$ K. 
This has an important effect on the energy production as well as on the 
resulting peak temperature achieved during the TNR. For instance, peak 
values for the nuclear energy generation rate and temperature achieved in Model 
ONe5, $\epsilon_{nuc,max} = 2.1 \times 10^{16}$ $\rm erg \, g^{-1} \, s^{-1}$ 
and $T_{max} = 2.44 \times 10^8$ K, translate into $\epsilon_{nuc,max} = 3.0 
\times 10^{16}$ $\rm erg \, g^{-1} \, s^{-1}$ and $T_{max} = 2.51 \times 10^8$ 
K, for Model ONe125B. A similar trend was also pointed out by Starrfield et al. 
(1998), using updated 
nuclear reaction rates with respect to Caughlan \& Fowler's (1988) ones.

Such differences have also some influence on the last phases of the evolution, 
when the envelope expands, cools down and eventually a fraction of the 
formerly accreted shells is ejected. In general, we find that models 
computed with 
updated rates lead to slightly larger ejected masses with larger mean kinetic 
energies (see Tables 4--6) as a result of the larger $\epsilon_{nuc,max}$ 
attained.  It is worth noticing that these differences in the 
properties of the TNRs will be reflected in the accompanying nucleosynthesis. 
The reason is twofold: first, the update of the reaction rates implies a 
certain modification of the nuclear paths. And second, the differences found 
in the time evolution as well as on the peak temperatures achieved modify the 
role played by (p,$\gamma$) and (p,$\alpha$) reactions.

\subsection{Nucleosynthesis in the NeNa-MgAl cycles}

 Due to the higher peak temperatures achieved in the models computed with the
 updated network (recommended rates. See Tables 4--6), ${}^{20}$Ne is slightly 
 reduced in the ejecta with respect to previous results obtained with older 
 rates (Jos\'e \& Hernanz 1998).  Differences are rather small 
 (i.e., a mean mass fraction of 0.17 by mass, instead of 0.18,
 results for the 1.15 \& 1.25 \msun models), but they turn out to be important 
 to determine the synthesis of less abundant species within the NeNa cycle. 
 For instance, the abundance of ${}^{21}$Ne increases by a factor between 
 $\sim 3-5$ (the higher temperatures as well as the lower rate adopted
 for $^{21}$Na(p,$\gamma$)$^{22}$Mg favor the chain 
 $^{20}$Ne(p,$\gamma$)$^{21}$Na($\beta^+$)$^{21}$Ne), 
 however no significant change is found for ${}^{22}$Ne (which is essentially
 reduced by (p,$\gamma$) reactions from its initial amount).
 Worth noticing is also the net increase in the final amount of ${}^{22}$Na 
 in the ejecta: a factor of $\sim 4-5$ for the 1.15 \& 1.25 \msun models
 (resulting from the lower rates adopted for 
  ${}^{21,22}$Na(p,$\gamma$)${}^{22,23}$Mg). This
 may have important consequences for gamma-ray astronomy, since it translates
 into a change by a factor of $\sim 2$ in the maximum expected distance 
 at which the 1275 keV ${}^{22}$Na line emitted by an 
 exploding ONe novae  would be eventually detected.
  Other isotopes, such as ${}^{23}$Na and ${}^{25}$Mg, are 
 also overproduced (except for the 1.35 \msun model), but less
 efficiently.  As a result, the isotopic ratios ${}^{23}$Na/${}^{22}$Na 
 decrease by nearly a factor of $\sim 2-3$.    
 The abundances of the other magnesium isotopes, ${}^{24,26}$Mg, are
 slightly enhanced (in some cases even by a factor of $\sim 2-4$). 
 Another interesting feature is that ${}^{26}$Al remains essentially
 unaffected by the update of the network 
  when the recomended rates are used. Therefore, the conclusions 
 relative to the small contribution of classical nova outbursts to
 the synthesis of the Galactic ${}^{26}$Al (see 
 Jos\'e, Hernanz \& Coc 1997) still hold. Since the amount of ${}^{27}$Al 
 (and that of ${}^{28}$Si) in the ejecta remains essentially
  unchanged, no variation on the isotopic ratio ${}^{26}$Al/${}^{27}$Al
 is found. 

  Another relevant outcome from the nucleosynthetic viewpoint 
is the dispersion in the mean ejected abundances obtained when
the uncertainties associated with the nuclear reaction rates are taken 
into account. In the following, we will analyse the impact of such
uncertainties in the resulting mean abundances in the ejecta. For that
purpose, we will compare the yields obtained when the two extreme networks
listed in Table 3 (i.e., A and C, leading to maximum and minimum
${}^{22}$Na-${}^{26}$Al production, respectively) are adopted. 
 We define the {\it dispersion factor}, $F$, as the ratio between the mean 
ejected abundances of a given nuclear species obtained when  network A 
 and C are adopted. 
 All models computed (i.e., involving 1.15, 1.25 or 1.35 \msun white
 dwarfs) show a very small dispersion in the ${}^{20}$Ne yields, 
 which results essentially from the different peak temperatures
 attained at the envelope rather than from uncertainties in the nuclear
 reactions (such as ${}^{20}$Ne(p,$\gamma$)${}^{21}$Na). 
 No relevant dispersion is found either for ${}^{22}$Ne, 
 ${}^{23}$Na, ${}^{27}$Al, and ${}^{28}$Si.
 On the contrary, the ${}^{21}$Ne yields show a rather wide dispersion: 
 a factor $F \sim 25$ in the 1.35 \msun models.
  Concerning ${}^{22}$Na, differences between $F \sim 2-4$ are found. 
 Similar degrees of dispersion are also obtained for the magnesium isotopes, 
 with $F$ ranging from 3 to 5 in the case of ${}^{25}$Mg and from 2 to 3 for
 ${}^{24}$Mg. Dispersion factors between 0.3 and 0.7  result  
 for ${}^{26}$Mg. The variation in the ejected amounts of ${}^{26}$Al is 
 particularly worth noticing, with a dispersion factor ranging from 
 $F \sim$ 4 to 7 (not considering the possible uncertainty associated
 with the 188 keV resonance in $^{26}$Al$^g$(p,$\gamma$)$^{27}$Si).
 Also interesting to notice is the fact that the 
${}^{26}$Al/${}^{27}$Al ratio remains in the range 0.5--0.1 (see Tables 4--6). 

 In view of the abovementioned analysis, the nuclear uncertainties
 accompanying the reaction rates within the NeNa-MgAl cycles introduce
 a relatively wide dispersion in the yields resulting from classical nova
 outbursts.
Therefore, predictions of ${}^{22}$Na and ${}^{26}$Al yields would 
benefit from new nuclear physics experiments aimed at reducing 
the uncertainties associated with some key reactions of the
NeNa-MgAl cycles, in particular \nampgmg and \altosi whose rates are
 uncertain by several orders of magnitude but also
$^{23}$Na(p,$\gamma)^{24}$Mg and \napgmg.
In addition, a verification of the yet unpublished values corresponding to the
0.188 MeV resonance of \algtosi, as measured by \cite{Vog89}, is also
recommended, due to its crucial role on the synthesis of ${}^{26}$Al.
 These experiments should be focused on the measurement of a few key
parameters (level energies, spectroscopic factors, ...). Many of these
measurements involve short lived radioactive species and should benefit
from the current development of radioactive ion beam facilities.

 On the contrary, our calculations have shown that the
remaining uncertainties on $^{22}$Mg(p,$\gamma)^{23}$Al,
$^{23}$Al(p,$\gamma)^{24}$Si, $^{23}$Mg(p,$\gamma)^{24}$Al,
$^{24}$Al(p,$\gamma)^{25}$Si, $^{26m}$Al(p,$\gamma)^{27}$Si,
$^{26}$Si(p,$\gamma)^{27}$P, $^{26}$Mg(p,$\gamma)^{27}$Al,
$^{27}$P(p,$\gamma)^{28}$S or $^{27}$Si(p,$\gamma)^{28}$P
have no effect on $^{22}$Na and $^{26}$Al production.

\section{Conclusions}

We have computed a series of hydrodynamic models of nova outbursts,
from the onset of accretion up to the ejection stage, for a range of ONe
white dwarfs with masses between 1.15 and 1.35 \msun, with the aim of
analyzing in detail the main nuclear paths leading to the 
synthesis of ${}^{22}$Na and ${}^{26}$Al. The role played by several
key reactions within the NeNa-MgAl cycles has been tested by a series
of models, which have been compared with previous results obtained by
Jos\'e \& Hernanz (1998) with different prescriptions for the reaction
rates. Limits on the production of both ${}^{22}$Na and ${}^{26}$Al 
have been derived
from a series of computations assuming upper, recommended or lower
estimates to the reaction rates.
The most relevant conclusions extracted from this work can be summarized
as follows:

1. The update of the nuclear reaction network results in a net
   increase in the final amount of ${}^{22}$Na ejected into the
   interstellar medium during classical nova outburts. This translates
   into an increase (by a factor of $\sim 2$) of the expected maximum distance 
   at which an exploding ONe nova would be eventually detected through its
   emission at  1275 keV (${}^{22}$Na decay $\gamma$-ray line). 

2. The final amount of ${}^{26}$Al remains essentially unaffected by
   the update of the network, confirming that classical novae scarcely 
   contribute to the Galactic ${}^{26}$Al,
   as pointed out earlier by Jos\'e, Hernanz \& Coc (1997).

3. Large nuclear uncertainties affect some key reactions of the NeNa-MgAl
   cycles, with a significant effect on the production of both ${}^{22}$Na
   and ${}^{26}$Al. When a combination of reaction rates, leading to
   maximum or minimum ${}^{22}$Na-${}^{26}$Al synthesis, is adopted,
   a large dispersion in the final abundances is found. 
   We stress that the derived ranges for ${}^{22}$Na-${}^{26}$Al production
   can be interpreted as error bars on sodium \& aluminum production
   posed by nuclear physics uncertainties.

4. In order to reduce the impact of the nuclear uncertainties in the production
   of ${}^{22}$Na and ${}^{26}$Al, we point out some nuclear reactions
   that deserve new experiments, in particular,
   \nampgmg,  \napgmg  and \altosi. 
   A confirmation of the values 
   corresponding to the 0.188 MeV resonance of \algtosi, as measured by 
   \cite{Vog89}, would also be of great importance.

\newpage
\appendix
\section{Appendix: Nuclear reactions}

We discuss here the reaction rates corresponding to proton capture reactions
on short-lived nuclei. Other reactions within the NeNa and MgAl cycles are
analyzed in the text (see Sections 3.2 and 4.2) and more extensively 
in the forthcoming {\em Compilation of Charged--Particle Induced 
Thermonuclear
Reaction Rates} (Angulo et al. 1998). First, we introduce some standard 
nuclear physics quantities and notations to be used hereafter.
 We also briefly summarize several indirect methods that are generally
used to extract or estimate unavailable direct data, for the calculation of
thermonuclear rates.

For an $A(x,y)B$ reaction, resonance strengths (at $E=E_r$) are given by
$\omega\gamma$, where $\omega=(2J_r+1)\left((2J_x+1)(2J_A+1)\right)^{-1}$
is the statistical spin factor and $\gamma=\Gamma_x\Gamma_y/\Gamma$ is the
width ratio.  The total width ($\Gamma(E)$) corresponds to the sum of 
partial widths
($\Gamma_I(E))$ : $\Gamma=\sum_{I=x,y,\ldots}\Gamma_I$.
The spectroscopic factor, $C^2S$, relates single particle widths 
($\Gamma^{s.p.}_x$) to actual ones:
$\Gamma_x$ = $C^2S\;\Gamma^{s.p.}_x$.
These single particle widths are calculated by solving the Schr\"odinger
equation for a particle in a realistic nuclear potential.
An approximation for the upper limit of single particle widths is given 
by the {\it Wigner limit}: 
$\Gamma^W_x=3\hbar^2({\mu}s^2)^{-1}P_{\ell_{x}}(E=E_r,s)$, where $s$ 
is the channel radius, $\mu$ the reduced mass, $P_{\ell_{x}}(E,s)$ the 
penetration factor and $\ell_x$ the orbital angular momentum transferred by 
particle $x$. We introduce the reduced width, $\theta^2_x$, defined by
$\Gamma_x = \theta^2_x\Gamma^W_x$, 
as it is sometimes used to derive estimates of
reaction rates (see Iliadis 1997 for a dicussion of the relationship between
$\theta^2$ and $C^2S$).

Experimental spectroscopic factors can be extracted from transfer reactions 
(e.g., the proton transfer reaction X($^3$He,d)Y will lead to one proton
spectroscopic factors used to calculate proton widths entering into the 
calculation of the X(p,$\gamma$)Y rate). Spectroscopic factors for conjugate 
reactions (obtained by the p $\rightleftharpoons$ n exchange) are assumed to 
be approximately equal. For instance, neutron spectroscopic factors for 
$^{22}$Ne can be extracted from the one neutron $^{21}$Ne(d,p)$^{22}$Ne 
transfer reaction. According to the above approximation, these {\it neutron}
spectroscopic factors can be used to determine {\it proton} widths for the 
calculation of the $^{21}$Na(p,$\gamma)^{22}$Mg rate for instance.
Radiative  widths ($\Gamma_\gamma$) corresponding to
transitions of the magnetic ($M\lambda$) or electric type ($E\lambda)$ of 
order $\lambda$ are often expressed in Weiskopf's units (W.u.) to remove the 
effect of their $E_\gamma^{2\lambda+1}$ dependence and strong variation with 
$\lambda$ (e.g. Firestone et al. 1996). Compilations of $\Gamma_\gamma$, 
expressed in 
these units, are available (Endt 1979) and may be used to estimate 
unknown radiative widths. They can also be obtained from the conjugate nuclei 
after correction for the $E_\gamma^{2\lambda+1}$ dependence. If particle 
emission is energetically forbbiden in the conjugate level, the radiative
width equals the total width.  
 If the lifetime of this level is known, the radiative width is
easily obtained. 

\subsection{The $^{21}$Na(p,$\gamma)^{22}$Mg reaction}

Estimates of the $^{21}$Na(p,$\gamma)^{22}$Mg reaction rate (\cite{CF88}) have 
been provided by \cite{WGTR86} and \cite{WL86}, considering the first three 
levels (Endt 1990) above the proton threshold. With the exception of the first 
level, the proton widths are much larger than the gamma widths, so that 
$\omega\gamma\approx\omega\Gamma_\gamma$. Only the first two levels 
($E_x$=5.714 and 5.837~MeV) contribute to the rate in the temperature domain 
considered.  The $E_x$ = 5.837~MeV level is assumed to be the conjugate of the 
$E_x$ = 5.910 one in $^{22}$Ne (Endt 1990). 
If this assignment is correct, the corresponding 
gamma width can be reliably deduced (Wiescher \& Langanke 1986).
On the contrary, the strength of the first $E_x$=5.714~MeV, $J^\pi$=2$^+$
level suffers from a significant uncertainty. 
For this level, the total width is known experimentally to be 
$\Gamma$=16.5$\pm$4.4~meV. To calculate the corresponding resonance strength, 
\cite{WL86} and \cite{WGTR86} estimated the proton width, $\Gamma_p$, and 
deduced the radiative width, $\Gamma_\gamma$, from the relation 
$\Gamma$ = $\Gamma_p$ + $\Gamma_\gamma$. To estimate the proton width, two 
 hypotheses were considered: \cite{WGTR86} assumed $\ell_p=0$ and 
$\theta^2_p=0.01$, but later \cite{WL86} took $\ell_p=2$ and 
$\theta^2_p=0.5$. Nevertheless, because of the $\Gamma_p$+
$\Gamma_\gamma$ = 16.5$\pm$4.4~meV constraint, 
the corresponding width ratios, $\gamma$, are 
not much different: 3.4 or 3.8~meV, 
very close to the maximum value ($\Gamma$/4), obtained when $\Gamma_\gamma$ =
$\Gamma_p$ = $\Gamma$/2. There are two reasons to think that, on the contrary,
$\Gamma_p\ll\Gamma\approx\Gamma_\gamma$ and that the corresponding strength is 
much smaller than the estimates provided by Wiescher et al. First, this level 
has a known counterpart in $^{22}$Ne at $E_x$=6.120~MeV (Endt 1998) with a
measured, purely radiative, total width $\Gamma$=29$\pm$9~meV, that
enables to calculate the
radiative width of the $E_x=5.714$ MeV $^{22}$Mg level to be
$\Gamma_\gamma\simeq$23~meV. Since it is found to be greater than the measured
total width (16.5$\pm$4.4~meV), it is likely that $\Gamma_p\ll\Gamma$ for this
level. The second indication comes from experimental data on the
$^{21}$Ne(d,p)$^{22}$Ne reaction, which provides information on neutron capture
on $^{21}$Ne (i.e., the mirror counterpart of proton capture on $^{21}$Na).
Neutron spectroscopic factors were obtained by \cite{Neo72} for many $^{22}$Ne
levels, from $E_x$=0 to 9.07~MeV, but not for the $E_x$=6.120~MeV one due to
its very low population and flat angular distribution. Indeed, the direct
comparison of peak heights in the experimental spectrum (fig.~1, Neogy et al.
1972) shows that, within levels of same spin and transferred angular momentum,
the cross section for the production of the $E_x$=6.120~MeV level is very
small. In addition, the corresponding angular distribution (fig.~5, Neogy et
al. 1972) is very flat, indicating no evidence for a direct component and hence
a very small neutron spectroscopic factor. This conclusion can be extended to
the proton spectroscopic factor in the $^{22}$Na mirror level.
The 6.551~MeV $^{22}$Na level, is assumed (Endt 1990,Endt 1998) to belong to
the same isospin triplet as the 6.120~MeV $^{22}$Ne level. Its proton
spectroscopic factor has been measured by Garrett et al. (1971) to be
$\approx$0.1. This value is not consistent with the Neogy et al. (1972) data
and would result, for the 5.714~MeV $^{22}$Mg level, in a proton width in
excess of the measured, 16.5~meV, total width. Hence one may question the
isospin triplet assignment made by Endt (1990) and Endt (1998). Moreover,
the 5.714~MeV $^{22}$Mg level is withdrawn from the triplet in the last
paper by Endt (1998).
 Accordingly, we 
consider the Wiescher et al. estimate for this resonance strength as an upper 
limit and adopt $\omega\gamma$ = 2.5, 0.25, 0.0~meV for upper 
($\Gamma_\gamma$ = $\Gamma_p$ = $\Gamma$/2), recommended value (with the
usual 0.1 reduction factor) and lower limit.
We provide below the updated formula for the rate including the uncertainty
factors and the direct capture contribution from Caughlan \& Fowler (1988):
\begin{eqnarray*}
1.41\times10^5\,T_9^{-3/2}\exp\left(-20.739/T_9^{1/3}-(T_9/.366)^2\right)
\times&\\
(1.+.02\,T_9^{1/3}+4.741\,T_9^{2/3}+ 667\,T_9+16.38\,T_9^{4/3}+
5.858\,T_9^{5/3})+& DC\;(CF88)\\
(0 \to 10.)\times40.8.\,T_9^{-3/2}\exp\left(-2.52\,/T_9\right)+& (E_x=5.714)\\
1857.\,T_9^{-3/2}\exp\left(-3.95\,/T_9\right)+ & (E_x=5.837)\\
(0.1 \to 10.)\times408.\,T_9^{-3/2}\exp\left(-5.49\,/T_9\right)&(E_x=5.965) \\
\end{eqnarray*}
The third term corresponds to the 5.837~MeV level, assuming that 
the analog assignment made by Endt (1990) is correct. This is however not 
granted as it is not present anymore in Endt (1998).    
The last term corresponds to the $E_x$=5.965~MeV, $J^\pi=0^+$ level. 
Since its widths are unknown (save that 
$\Gamma_\gamma\ll\Gamma\approx\Gamma_p$), \cite{WGTR86} assumed a 
typical value for $\Gamma_\gamma$ based on the statistics of Endt (1979).
We introduce a factor of ten uncertainty to account for the dispersion of 
the gamma strengths within this statistics, but it has a little 
influence on the rate. The corresponding rates, relative to Caughlan \& Fowler 
(1988) (i.e., Wiescher et al. 1986) are shown in Figure~4: the maximum effect 
occurs between $T_8$ = 0.5 and 3, well within the nova temperature range.
Experimental data of the $E_x$=5.714~MeV, $J^\pi$=2$^+$ level is clearly 
needed to improve the reliability of this rate. 

\subsection{The $^{22}$Mg(p,$\gamma)^{23}$Al and 
$^{23}$Al(p,$\gamma)^{24}$Si reactions}

 As ${}^{22}$Mg plays a crucial role on the synthesis of ${}^{22}$Na 
through its beta decay, its destruction by 
$^{22}$Mg(p,$\gamma)^{23}$Al has to be considered.
The contribution of the direct capture to 
the ground state for the $^{22}$Mg(p,$\gamma)^{23}$Al rate has been calculated 
by \cite{WGTR86}. Later on, \cite{WGS88} measured the location of the first 
resonance and also calculated its strength (shell model and mirror level). 
But due to the very low Q--value (0.125~MeV), rapid photodisintegration of 
$^{23}$Al prevents $^{22}$Mg destruction. Another destruction channel has also 
been proposed: it takes place via two proton captures on $^{22}$Mg (i.e., 
proton capture on the small population of $^{23}$Al through 
$^{23}$Al(p,$\gamma)^{24}$Si. G\"orres, Wiescher \& Thielemann 1995)
and a new rate was proposed by \cite{Her95}, based on shell model
calculations. 
The location of the first two $^{24}$Si levels has been recently determined 
experimentally (Schatz et al.  1997): it leads to a higher rate than in 
\cite{Her95} due to the lower location of the first resonance 
($E_r^{cm}=0.141\pm0.031$~MeV, instead of 320~keV). 
In our nova models, we have  adopted the new limits for the 
$^{23}$Al(p,$\gamma)^{24}$Si rate given by \cite{Sch97}. 

\subsection{The $^{23}$Mg(p,$\gamma)^{24}$Al and $^{24}$Al(p,$\gamma)^{25}$Si 
            reactions}

%====mg23pg====
An estimate to the $^{23}$Mg(p,$\gamma)^{24}$Al rate has been provided by
\cite{WGTR86} based on direct capture and three resonances.
%corresponding to two known $^{24}$Al levels plus one inferred from $^{24}$Na. 
Since then, Kubono, 
Kajino \& Kato (1995), Endt (1998) and Herndl et al. (1998) have discussed 
analog level assignments, in particular with the help of shell model 
calculations and isobaric multiplet mass equation (Herndl et al. 1998). 
For nova 
nucleosynthesis, only the two first resonances have to be considered. They 
correspond to the $E_x$ = 2.349 and 2.534~MeV levels and are assumed to be the 
analogs of the 2.514~MeV, 3$^+$ and 2.563~MeV, 4$^+$(2$^+$) ones in $^{24}$Na 
(Endt 1998, Herndl et al. 1998).
 In addition to the estimate provided by \cite{WGTR86} the strength of the
second resonance has been calculated (shell model) by \cite{Her98}.  
However, combining  statistics of reduced radiative widths (Endt 1979) 
together with the spectroscopic factor reported in \cite{End78}, show
that its contribution is, in any case, negligible with respect to the 
first one in the domain of temperature considered here.
The strength of the first resonance
is deduced from the analog level ($\Gamma_\gamma\ll\Gamma_p$) and 
hence suffers only small uncertainty like the direct capture contribution (see 
Wiescher et al. 1986). 
In the domain of nova nucleosynthesis, the main 
uncertainty comes from the determination of the energy of the first level: 
taking $E_r^{cm}$ = 0.458 (Kubono et al. 1995), 0.478 (Endt 1998; Herndl et al.
1998) or 0.51~MeV (Wiescher et al. 1986)
 results in a variation of the rate of less than a factor of 
$\sim$ 10 around $T_8$ = 2.5. 
To check the importance of this reaction, we adopted the highest rate (Kubono
et al. 1995) in our test calculations.

%====al24pg====
For $^{24}$Al(p,$\gamma)^{25}$Si, we use the rate proposed by \cite{Her95},
based on shell model calculations. It remains very uncertain, due to the 
limited experimental spectroscopic information on $^{25}$Si. It should have, 
however, very little effect on Ne--Na leaks, since either 
$^{24}$Al$(\beta^+)^{24}$Mg(p,$\gamma)^{25}$Al or
$^{24}$Al(p,$\gamma)^{25}$Si($\beta^+)^{25}$Al lead to $^{25}$Al in the MgAl 
cycle.

\subsection{The $^{25}$Al(p,$\gamma)^{26}$Si reaction}

$^{25}$Al(p,$\gamma)^{26}$Si is important since it leads to the formation
of the short-lived isomer ($^{26}$Al$^m$) instead of the long-lived ground
state ($^{26}$Al$^g$).
Its rate suffers from large uncertainties in the domain of nova
nucleosynthesis due mainly to the unknown location 
of the analog of the $^{26}$Mg $E_x$ = 6.13~MeV, 3$^+$ level. A level shift
similar to the one of its immediate neighbor ($E_x$ = 6.26~MeV; 0$^+$ in
$^{26}$Mg) would bring the $3^+$ level within the Gamow peak with a dramatic
influence on the rate.
The maximum effect on $^{26}$Al production is found when
$E_r^{cm}(3^+) = 0.2\pm0.1$~MeV (Coc et al. 1995).
New Coulomb displacement energy calculations (Iliadis et al. 1996) give
$E_r^{cm}(3^+) = 0.45\pm$0.1~MeV.  However, discrepancies
 between calculated and experimentally known $^{26}$Si
level energies can exceed $\pm$0.1~MeV so that the case
$E_r^{cm}(3^+) = 0.2$ cannot be ruled out.
(See, e.g., the $^{23}$Al(p,$\gamma)^{24}$Si case above where the location
of the first resonance is found $\simeq$180~keV below the calculated one.)
 Accordingly, for this rate, we used the lower limit, recommended value
and higher limit provided by Coc et al. (Case A, B and C, respectively).     

\subsection{The $^{26}$Si(p,$\gamma)^{27}$P reaction}

The $^{26}$Si(p,$\gamma)^{27}$P reaction rate has been discussed by 
\cite{WGTR86}, and more recently by \cite{Her95}. Only two excited states are
known in $^{27}$P and only one ($E_x$ = 1.660~MeV, $J^\pi$ = $5/2^+$) is low
enough to be of astrophysical interest. From the comparison with the spectrum
of the mirror nucleus, another level, $J^\pi$ = $3/2^+$, located at $E_x$ =
0.985~MeV in $^{27}$Mg, is expected below the $5/2^+$ one. \cite{WGTR86}
assumed a small shift, while shell model calculations (Herndl et al. 1995) 
led
to a much larger shift, which resulted in a dramatic increase of the rate by
almost four orders of magnitude.
The calculated strength (Herndl et al. 1995) is in good agreement with the
$\Gamma_\gamma$ value (0.8~meV) inferred from the lifetime of the $E_x$ =
0.985, $^{27}$Mg,  conjugate level and with
the $\Gamma_p$ obtained from the neutron spectroscopic factor
in the conjugate $^{27}$Mg level reported in \cite{End78}.
In consequence, the uncertainty results from the unknown position of the
resonance. 
However, it does not affect strongly the rate (a factor of ten at
most) for $1.5<T_8<4$ down to $E_r^{cm}\lap$0.2~MeV due to the compensating
effect of the evolution of the penetrability $P_{l_{p}}(E_r,s)$ and
exp($-E_r/kT$) factors with $E_r$ small variations.
 The strength of the $5/2^+$ resonance is known from the \cite{WGTR86}
estimates, from $^{27}$Mg spectroscopy and from the \cite{Her95}
calculations. 
However, its location ($E_r$ = 0.76~MeV) makes its 
contribution negligible compared to the direct capture one at nova 
temperatures.  
The latter provides the lower limit for the rate, 
reflecting the possibility that the $J^\pi$ = $3/2^+$ resonance 
energy is lower than $E_r^{cm}\simeq$0.2~MeV, as considered by 
\cite{WGTR86}, while the upper limit is given by the \cite{Her95} rate.
One should note that this effect could be 
enhanced if the uncertain $^{25}$Al(p,$\gamma)^{26}$Si rate is close to its 
maximum value (see above).

\subsection{The $^{27}$P(p,$\gamma)^{28}$S, and 
$^{27}$Si(p,$\gamma)^{28}$P reactions}

%====p27pg====
Only the ground state of $^{28}$S is known but from the level scheme of the
conjugate nucleus, $^{28}$Mg, one can infer that no resonance is expected 
for $E_r^{cm} \lap $1.4~MeV. 
Hence, at the temperatures considered in the text, the 
$^{27}$P(p,$\gamma)^{28}$S reaction proceeds only through direct capture. The 
corresponding rate has been obtained by \cite{Her95} from shell model 
calculations. It is about a factor of ten higher than the Caughlan \&
Fowler (1988) one and 
can be considered as reliable enough (i.e. within an order of magnitude), due 
to the expected absence of resonances. 

%====si27pg====
The $^{27}$Si(p,$\gamma)^{28}$P reaction rate given in Caughlan \& Fowler
(1998) comes from the analysis of \cite{WGTR86}. They considered five 
resonances corresponding to the $^{28}$P levels between $E_x$ = 2.143 and 
2.628~MeV, out of which, for three levels, the spins are uncertain. From the 
assumed conjugate levels in $^{28}$Al, they extracted radiative widths and 
proton widths using spectroscopic factors obtained by neutron transfer on 
$^{27}$Al. More recently, the good agreement between calculated (Endt \& 
Booten 1993) and experimental (Endt 1990) level schemes, 
spectroscopic factors and radiative widths, for $A=28$ nuclei, 
 gives confidence to the $^{28}$Al 
and $^{28}$P conjugate level assignment (see also \cite{Ili98}). The first 
level above proton threshold, $E_x$ = 2.104~MeV, $J^\pi$ = 2$^+$, was not 
considered by \cite{WGTR86}, but its estimated proton width is however 
too small 
($\Gamma_p\lap\Gamma^W_p\simeq 10^{-20}$~eV) to contribute to the rate for 
$T_8 \gap 0.3$. The uncertainty remains limited to transposition of $^{28}$Al 
data to $^{28}$P. 

\acknowledgments
The authors are grateful to Jean-Pierre Thibaud for helpful and stimulating
discussions. We also thank Carmen Angulo for allowing us to use information 
from the NACRE compilation prior to publication, and the referee, 
Sumner Starrfield, for a critical reading of the manuscript.
This research has been partially supported by the CICYT-P.N.I.E. 
(ESP98-1348), by the DGICYT (PB97-0983-C03-02; PB97-0983-C03-03), by the 
CIRIT (GRQ94-8001), by the AIHF1997-0087, by the CHRXCT930339 and by the 
PICS 319.

\newpage

\begin{deluxetable}{cc}
\tablewidth{0 pt}
\tablecaption{Initial composition of the envelope (up to Si), assuming
 50\% mixing with the ONe white dwarf core}
\tablehead{
\colhead{Nuclei}&
\colhead{Mass fraction}
}
\startdata
 $^{1}$H   &3.5E-1 \nl
 $^{3}$He  &1.5E-5 \nl
 $^{4}$He  &1.4E-1 \nl
 $^{6}$Li  &3.2E-10\nl
 $^{7}$Li  &4.7E-9 \nl
 $^{9}$Be  &8.3E-11\nl
 $^{10}$B  &5.3E-10\nl
 $^{11}$B  &2.4E-9 \nl
 $^{12}$C  &6.1E-3 \nl
 $^{13}$C  &1.8E-5 \nl
 $^{14}$N  &5.5E-4 \nl
 $^{15}$N  &2.2E-6 \nl
 $^{16}$O  &2.6E-1 \nl
 $^{17}$O  &1.9E-6 \nl
 $^{18}$O  &1.1E-5 \nl
 $^{19}$F  &2.0E-7 \nl
 $^{20}$Ne &1.6E-1 \nl
 $^{21}$Ne &3.0E-3 \nl
 $^{22}$Ne &2.2E-3 \nl
 $^{23}$Na &3.2E-2 \nl
 $^{24}$Mg &2.8E-2 \nl
 $^{25}$Mg &7.9E-3 \nl
 $^{26}$Mg &5.0E-3 \nl
 $^{27}$Al &5.4E-3 \nl
 $^{28}$Si &3.3E-4 \nl
\enddata
\end{deluxetable}

\newpage

\begin{deluxetable}{lccll}
\tablewidth{0 pt}
\tablecaption{Test models}
\tablehead{
\colhead{Nuclear reaction}& 
\colhead{Old rate}&
\colhead{Test rate}&
\colhead{$\rm \frac{X(^{22}Na)_{test}}{X(^{22}Na)_{old}}$}&
\colhead{$\rm \frac{X(^{26}Al)_{test}}{X(^{26}Al)_{old}}$}
}
\startdata
 1.15 \msun ONe    &     &     &     &     \nl
\cline{1-5}
\nampgmg & CF88  & CF88/100   &2.3 & 1.1 \nl
\altosi  & Wie86 & Coc95, case A& 1& 1.2 \nl
         & Wie86 & Coc95, case C& 1& 0.5 \nl
\natomg  & CF88  & CF88+GWR89 &1.2 & 1.3 \nl
\algtosi & Vog89 & Vog89      & 1  & 1.9 \nl
  &    & 1/3$\times$res(0.188 MeV) &    &     \nl
\almtosi & CF88  & CF88$\times$100   & 1  &  1  \nl
\sitop   & Wie86 & Her95      & 1  &  1  \nl
\mgtoal  & Ili90 & Ili90+Cha90& 1  &  1  \nl
\cline{1-5}
 1.25 \msun ONe    &     &     &     &     \nl
\cline{1-5}
\nampgmg & CF88  & CF88/100   & 3  & 1.1 \nl
\napgmg  & CF88  & Ste96      & 3  & 1.2 \nl
\mgpgal  & Wie86 & KKK95      &1.1 & 1.1 \nl
\cline{1-5}
 1.35 \msun ONe    &     &     &     &     \nl
\cline{1-5}
\nampgmg & CF88  & CF88/100   &1.2 &  1  \nl
\alsmall & Wor94 & Sch97      & 1  &  1  \nl
\enddata
\tablenotetext{}{References to the reaction rates: 
Wie86 (Wiescher et al. 1986),
CF88 (Caughlan \& Fowler 1988), GWR89 (G\"orres, Wiescher \& 
Thielemann 1989), Vog89 (Vogelaar 1989), 
Cha90 (Champagne et al. 1990), Ili90 (Iliadis et al. 1990),
Wor94 (van Wormer et al. 1994),
Coc95 (Coc et al. 1995), Her95 (Herndl et al. 1995), 
KKK95 (Kubono, Kajino \& Kato 1995),  
Ste96 (Stegm\"uller et al. 1996), Sch97 (Schatz et al. 1997)} 
\end{deluxetable}

\newpage
\begin{deluxetable}{lcccc}
\footnotesize
\tablecaption{Reaction rates adopted for the calculation of Maximum (A), 
  Recommended (B), and Minimum (C) ${}^{22}$Na-${}^{26}$Al production
  compared with those adopted in Jos\'e \& Hernanz (1998)} 
\tablewidth{0 pt}
\tablehead{
\colhead{Reaction}&
\colhead{JH98}&
\colhead{A}&
\colhead{B}&
\colhead{C}
}
\startdata
$^{21}$Na(p,$\gamma$)$^{22}$Mg& CF88   & This work  & This work  & This work\nl
                              &     & Lower limit& Recommended & Upper limit\nl
\cline{1-5}
$^{21}$Ne(p,$\gamma$)$^{22}$Na& CF88   & CF88       & CF88       & CF88     \nl
                           & with f=0.1& f=0.01     &f=0.001     &f=0       \nl
\cline{1-5}
$^{22}$Na(p,$\gamma$)$^{23}$Mg& CF88   & Ste96      & NACRE      & Ste96    \nl
                          &        & Lower limit & Recommended& Upper limit \nl
\cline{1-5}
$^{25}$Al(p,$\gamma$)$^{26}$Si& Wie86  & Coc95      & Coc95      & Coc95    \nl
                   &  & Case A (Lower)& Case B (Recommended) &Case C (Upper)\nl
\cline{1-5}
$^{23}$Na(p,$\gamma$)$^{24}$Mg& CF88   & CF88+GWR89 &CF88+GWR89  &CF88+GWR89\nl
                              &        &with f=1    &f=0.1       &f=0       \nl
\cline{1-5}
$^{23}$Mg(p,$\gamma$)$^{24}$Al& Wie86  & KKK95      &KKK95       &KKK95     \nl
\cline{1-5}
$^{26}$Mg(p,$\gamma$)$^{27}$Al& Ili90  &Cha90+Ili90 &Cha90+Ili90 &Cha90+Ili90\nl
                              &        &with f=1    &f=0.1       &f=0       \nl
\cline{1-5}
$^{26}$Al$^m$(p,$\gamma$)$^{27}$Si& CF88 &CF88      & CF88       & CF88     \nl
\cline{1-5}
$^{27}$Al(p,$\alpha$)$^{24}$Mg& Cha88  &Cha88+Tim88 &Cha88+Tim88 &Cha88+Tim88\nl
                              &        &with f=1    &f=0.1       &f=0       \nl
\cline{1-5}
$^{23}$Al(p,$\gamma$)$^{24}$Si& WW80   & Sch97      & Sch97      & Sch97    \nl
                              &        &Lower limit &Recommended &Upper limit\nl
\cline{1-5}
$^{26}$Si(p,$\gamma$)$^{27}$P & Wie86  & Her95      & Her95      & Her95    \nl
\cline{1-5}
$^{26}$Al$^g$(p,$\gamma$)$^{27}$Si& Vog89& Coc95    & Coc95      & Coc95    \nl
                              &        &with f1=f2=0& f1=f2=0.1  &f1=f2=1   \nl
\enddata
\tablenotetext{}{Additional references to the reaction rates: 
WW80 (Wallace \& Woosley 1980), Cha88 (Champagne et al. 1988), 
Tim88 (Timmermann et al. 1988), NACRE (Angulo et al. 1998). f, f1 and f2
represent uncertainty factors in the analytic rates.}

\end{deluxetable}

\newpage

\begin{deluxetable}{ccccc}
\tablewidth{0 pt}
\tablecaption{Main characteristics of the 1.15 \msun ONe nova models 
compared with model ONe3 (Jos\'e \& Hernanz 1998)}
\tablehead{
\colhead{}&
\colhead{ONe3}&
\colhead{ONe115A}&
\colhead{ONe115B}&
\colhead{ONe115C}
}
\startdata
Reaction Network           & Old&  A  &  B  & C   \nl
$\Delta M_{env}$ ($10^{-5}$ \msun)
                              & 3.2 & 3.2 & 3.2 & 3.2 \nl
$T_{max}$ ($10^8$ K)          & 2.19& 2.33& 2.31& 2.30 \nl
$\Delta M_{ejec}$ ($10^{-5}$ \msun) 
                              & 1.9 & 2.6 & 2.6 & 2.6  \nl
$K$ ($10^{45}$ erg)           & 1.2 & 1.5 & 1.5 & 1.5 \nl
 $^{20}$Ne &1.8E-1 & 1.6E-1 & 1.7E-1 & 1.8E-1 \nl
 $^{21}$Ne &3.0E-5 & 2.0E-4 & 8.7E-5 & 2.5E-5 \nl
 $^{22}$Ne &1.7E-3 & 1.6E-3 & 1.5E-3 & 1.5E-3 \nl
 $^{22}$Na &5.3E-5 & 5.3E-4 & 2.7E-4 & 1.2E-4 \nl
 $^{23}$Na &7.5E-4 & 1.3E-3 & 1.4E-3 & 1.4E-3 \nl
 $^{24}$Mg &1.0E-4 & 9.6E-4 & 4.1E-4 & 3.5E-4 \nl
 $^{25}$Mg &2.9E-3 & 6.1E-3 & 3.9E-3 & 1.8E-3 \nl
 $^{26}$Mg &3.4E-4 & 8.3E-4 & 6.4E-4 & 1.2E-3 \nl
 $^{26}$Al &9.3E-4 & 1.2E-3 & 8.0E-4 & 3.3E-4 \nl
 $^{27}$Al &4.5E-3 & 4.8E-3 & 3.5E-3 & 3.4E-3 \nl
 $^{28}$Si &5.4E-2 & 6.8E-2 & 5.8E-2 & 5.6E-2 \nl
\enddata
\end{deluxetable}

\newpage

\begin{deluxetable}{ccccc}
\tablewidth{0 pt}
\tablecaption{Main characteristics of the 1.25 \msun ONe nova models 
compared with model ONe5 (Jos\'e \& Hernanz 1998)}
\tablehead{
\colhead{}&
\colhead{ONe5}&
\colhead{ONe125A}&
\colhead{ONe125B}&
\colhead{ONe125C}
}
\startdata
Reaction Network           & Old &  A  &  B  & C   \nl
$\Delta M_{env}$ ($10^{-5}$ \msun)
                              & 2.2 & 2.2 & 2.2 & 2.2 \nl
$T_{max}$ ($10^8$ K)          & 2.44& 2.54& 2.51& 2.51\nl
$\Delta M_{ejec}$ ($10^{-5}$ \msun) 
                              & 1.4 & 1.8 & 1.8 & 1.8 \nl
$K$ ($10^{45}$ erg)           & 1.4 & 1.5 & 1.5 & 1.5 \nl
 $^{20}$Ne &1.8E-1 & 1.6E-1 & 1.7E-1 & 1.7E-1 \nl
 $^{21}$Ne &3.5E-5 & 3.7E-4 & 1.4E-4 & 3.0E-5 \nl
 $^{22}$Ne &1.0E-3 & 9.8E-4 & 9.1E-4 & 8.6E-4 \nl
 $^{22}$Na &9.6E-5 & 6.9E-4 & 3.5E-4 & 1.9E-4 \nl
 $^{23}$Na &1.4E-3 & 1.6E-3 & 1.9E-3 & 1.9E-3 \nl
 $^{24}$Mg &2.0E-4 & 9.0E-4 & 4.1E-4 & 3.9E-4 \nl
 $^{25}$Mg &2.4E-3 & 4.6E-3 & 3.2E-3 & 1.3E-3 \nl
 $^{26}$Mg &2.8E-4 & 5.9E-4 & 5.1E-4 & 1.1E-3 \nl
 $^{26}$Al &5.4E-4 & 8.8E-4 & 6.5E-4 & 2.0E-4 \nl
 $^{27}$Al &2.0E-3 & 2.9E-3 & 2.2E-3 & 2.2E-3 \nl
 $^{28}$Si &5.6E-2 & 6.8E-2 & 5.8E-2 & 5.6E-2 \nl
\enddata
\end{deluxetable}

\newpage

\begin{deluxetable}{ccccc}
\tablewidth{0 pt}
\tablecaption{Main characteristics of the 1.35 \msun ONe nova models 
compared with model ONe6 (Jos\'e \& Hernanz 1998)}
\tablehead{
\colhead{}&
\colhead{ONe6}&
\colhead{ONe135A}&
\colhead{ONe135B}&
\colhead{ONe135C}
}
\startdata
Reaction Network           & Old &  A  &  B  & C   \nl
$\Delta M_{env}$ ($10^{-5}$ \msun)
                              & 0.54 & 0.54& 0.54 & 0.54 \nl
$T_{max}$ ($10^8$ K)          & 3.24 & 3.22& 3.31 & 3.23 \nl
$\Delta M_{ejec}$ ($10^{-5}$ \msun) 
                              & 0.44& 0.44 & 0.44 & 0.44 \nl
$K$ ($10^{45}$ erg)           & 0.9 & 1.0  & 1.0  & 1.0  \nl
 $^{20}$Ne &1.5E-1 & 1.3E-1 &  1.4E-1 & 1.4E-1 \nl
 $^{21}$Ne &5.1E-5 & 9.9E-4 &  2.6E-4 & 3.9E-5 \nl
 $^{22}$Ne &1.5E-4 & 1.9E-4 &  1.9E-4 & 1.7E-4 \nl
 $^{22}$Na &6.0E-4 & 1.4E-3 &  1.0E-3 & 7.7E-4 \nl
 $^{23}$Na &6.6E-3 & 5.2E-3 &  5.5E-3 & 5.6E-3 \nl
 $^{24}$Mg &3.6E-4 & 8.6E-4 &  5.4E-4 & 5.4E-4 \nl
 $^{25}$Mg &4.2E-3 & 6.8E-3 &  4.5E-3 & 1.3E-3 \nl
 $^{26}$Mg &5.9E-4 & 6.3E-4 &  7.2E-4 & 2.1E-3 \nl
 $^{26}$Al &7.2E-4 & 1.1E-3 &  7.2E-4 & 1.5E-4 \nl
 $^{27}$Al &1.8E-3 & 2.1E-3 &  1.8E-3 & 2.0E-3 \nl
 $^{28}$Si &3.5E-2 & 5.0E-2 &  4.2E-2 & 4.1E-2 \nl
\enddata
\end{deluxetable}

\clearpage

\figcaption[jjosefig1.eps]{
   Snapshots of the evolution of $^{20,21,22}$Ne, $^{21,22,23}$Na,
   and $^{22,23}$Mg (in mass fractions) along the accreted envelope
   for a 1.25 \msun ONe novae accreting at a rate 
   $\dot{\rm{M}}$ = 2 \power{-10} \msyr. 
   The mass coordinate represents the mass below the surface, relative
   to the total mass. 
   The successive panels, from top to bottom, correspond to
   the time for which the temperature at the burning shell reaches
   5 \power{7}, 7 \power{7}, 10$^{8}$, 2 \power{8}, 
   $\rm{T_{max}} =$ 2.44 \power{8}, plus 3 panels corresponding to
   the last phases of the explosion, when the white dwarf envelope 
   has already expanded to a size of $\rm{R_{wd}} \sim 10^9$, $10^{10}$
   and $10^{12}$ cm, respectively. 
   The arrow indicates the base of the ejected shells.} 

\figcaption[jjosefig2.eps]{
   Main nuclear paths in both NeNa and MgAl cycles.}

\figcaption[jjosefig3.eps]{
   Same as Fig. 1, for the evolution of $^{24,25,26}$Mg, 
   $^{26}$Al$^{g,m}$, $^{25,27}$Al, and $^{26,27}$Si.}

\figcaption[jjosefig4.eps]{
   Comparison between the nuclear reaction rate for 
   $^{21}$Na(p,$\gamma$)$^{22}$Mg provided in this work, with the
   analytic rate given in Caughlan \& Fowler (1988).}

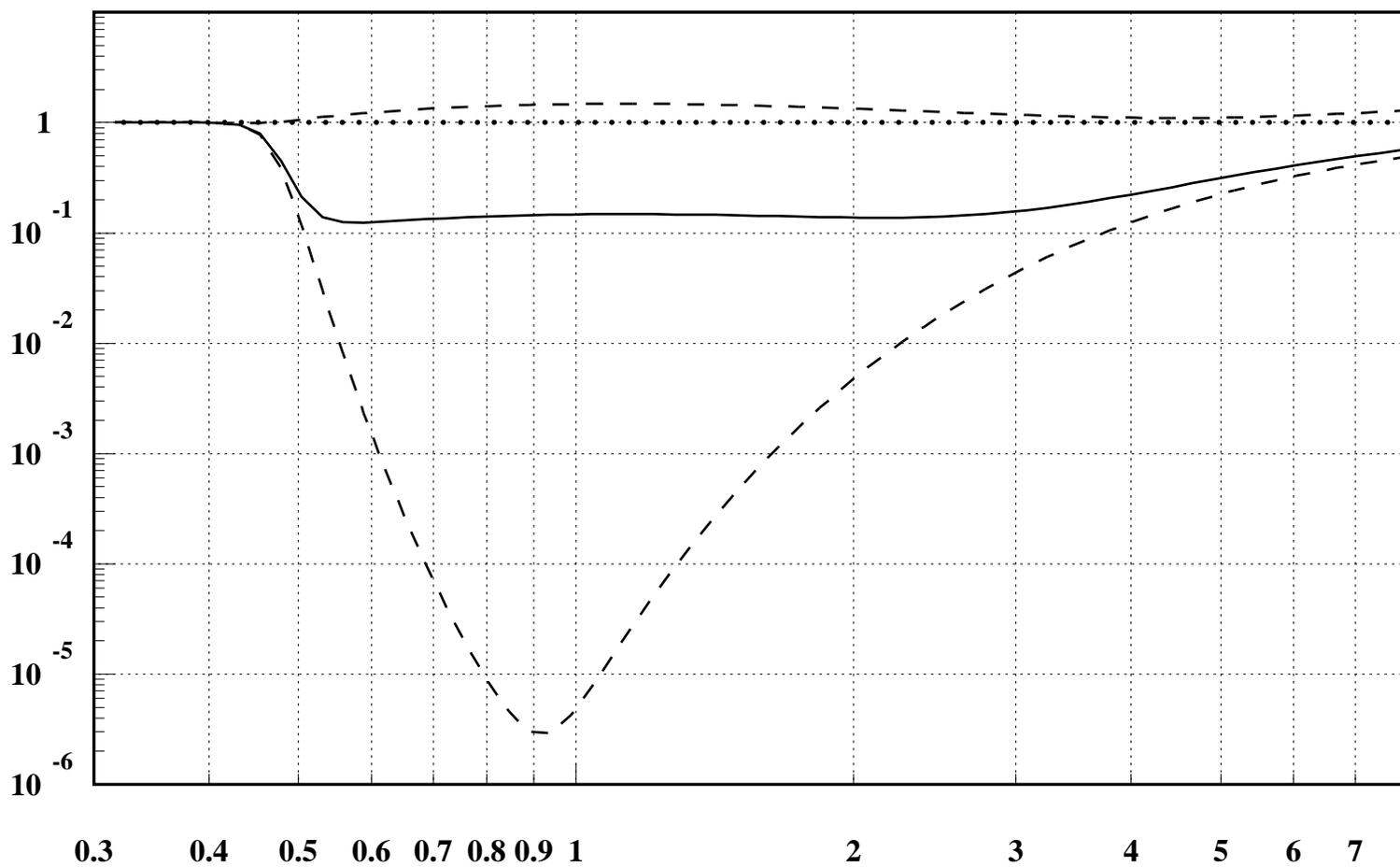
\end{document}